\documentstyle[prl,twocolumn,aps]{revtex}
\begin{document}
\draft
\title{Luttinger liquid behavior in multi-wall carbon nanotubes}
\author{Reinhold Egger}
\address{Fakult\"at f\"ur Physik, Albert-Ludwigs-Universit\"at, 
Hermann-Herder-Stra{\ss}e 3, D-79104 Freiburg, Germany}
\date{Date: \today}
\maketitle
\begin{abstract}
The low-energy theory for multi-wall carbon nanotubes including the
long-ranged Coulomb interactions, internal screening effects,
and single-electron hopping between
graphite shells is derived and analyzed by bosonization methods. 
Characteristic Luttinger liquid power laws are found for the
tunneling density of states,
with exponents approaching their Fermi liquid value 
only very slowly as the number of conducting shells increases. 
With minor modifications, the same conclusions apply to transport
in ropes of single-wall nanotubes.
\end{abstract}

\narrowtext
Metallic carbon nanotubes constitute a novel and exciting realization 
of one-dimensional (1D) conductors \cite{ijima91,tans97}.
The strong electronic correlations 
observed experimentally \cite{tans98}
 pose many challenges to theorists.
It is well known that electron-electron interactions invalidate the Fermi liquid
description in one dimension.  Often 1D conductors can instead 
 be described as a Luttinger liquid (LL) at low energy scales.
 The theoretical prediction \cite{egger97,kane97}
of LL behavior in a single-wall nanotube (SWNT) has indeed 
been verified in a recent transport experiment \cite{bockrath99}.
Furthermore, pronounced interaction effects have been observed
in experiments on individual multi-wall nanotubes (MWNT)
composed of several (typically 10 to 20) concentric graphite shells 
\cite{schoenen99,kasumov97}, most notably a pronounced zero-bias anomaly at 
low applied voltage, which was tentatively interpreted in terms of 
LL theory.  Since MWNTs are much easier to manipulate than SWNTs, 
interaction effects might be useful for 
applications, and whether a MWNT can display LL behavior or not could then 
be a question of practical importance.
Obviously, very similar questions [and answers given below]
apply to bundles (ropes) of SWNTs, the system actually 
studied in Ref.~\cite{bockrath99}.

 In this Letter, the low-energy theory of an individual MWNT 
composed of $N$ metallic graphite shells
\cite{foot0} with radii $R_1<R_2<\cdots<R_N$ is derived,
fully taking into account the externally unscreened Coulomb interaction, 
internal screening effects, and inter-shell electron tunneling. 
Insulating shells and the substrate are incorporated 
in terms of a (space-dependent) dielectric constant.  
The theory holds for energy scales $k_B T, eV \ll v/R_N$ [we set $\hbar=1$], where
$v\approx 8\times 10^5$ m/sec is the radius-independent Fermi velocity; for
 $R_N = 10$~nm,  $v/R_N \approx 350$~meV.  Then
 one needs to take into account only two transport bands $\alpha=\pm$  per shell.
For both bands, there is a right- and a left-moving $(r=\pm=R/L$)
branch with linear (massless) dispersion.   Regarding the importance of disorder,
both theory \cite{todorov98} and experiment 
\cite{tans97,schoenen99,frank98} are consistent with large 
elastic mean free paths (several hundred nm up to a few $\mu$m)
in metallic tubes.   Therefore we mainly focus on the clean case here.
By employing bosonization methods \cite{book}, 
pronounced LL effects are predicted for a MWNT, 
with only a slow crossover to Fermi liquid behavior as
$N$ increases. The precise relation between experimentally
 measurable exponents and
microscopic quantities turns out to be quite
different from the SWNT case discussed in Refs.~\cite{egger97,kane97}.

The construction of the low-energy theory starts by expanding the 
electron operator for spin
$\sigma=\pm$ on shell $n=1,\ldots,N$ in terms of the right- and 
left-moving Bloch waves $\phi_{r\alpha n}(x,y)$
for each  transport band $\alpha=\pm$,
\begin{equation} \label{expa}
\Psi_{\sigma n}(x,y) = \sum_{r\alpha} \phi_{r\alpha n}(x,y) \psi_{r\alpha\sigma n}(x)
\;.
\end{equation}
The Bloch waves are stiff around the circumference, $\phi_{r\alpha n}(x,y)=
(2\pi R_n)^{-1/2} \exp(-i\alpha k_F x)$, where $k_F=2\pi/3 a$ with the
lattice constant $a=2.46$~\AA,
 $0<y<2\pi R_n$, and $x$ is the transport direction.
The expansion  (\ref{expa}) allows to 
formulate the theory in terms of 1D fermion operators
$\psi_{r\alpha\sigma n}(x)$.   Ignoring  interactions 
and inter-shell tunneling for the moment, the linear dispersion
implies $N$ copies of a 1D massless Dirac Hamiltonian,
\begin{equation}\label{h0}
H_0= -i v \sum_{r\alpha\sigma n} r \, \psi^\dagger_{r\alpha\sigma n} \partial_x
\psi^{}_{r\alpha\sigma n} \;.
\end{equation}

In the next step, let us include the Coulomb interactions among the electrons.
Short-ranged backscattering processes are ignored since the corresponding
couplings are extremely
small even in a SWNT and scale as $1/R_n$ \cite{egger97}.
Since one is normally off half-filling due to the presence of metallic 
gates, Umklapp scattering is also neglected.  Under the expansion (\ref{expa}),
the important forward scattering processes lead to the contribution
\begin{equation} \label{hi} 
H_I=\frac12 \sum_{n,m=1}^N \int dx dx' \, \rho_n(x) V_{nm}(x-x') \rho_m(x')\;,
\end{equation}
where the 1D densities for shell $n$ are 
\begin{equation} \label{d1d}
\rho_n(x) = \sum_{r\alpha\sigma} \psi^\dagger_{r\alpha\sigma n}(x) 
\psi^{}_{r\alpha\sigma n}(x) \;.
\end{equation}
The effective 1D potential $V_{nm}=V_{mn}$ is obtained from the externally 
unscreened 3D Coulomb potential
by integrating over the circumferential coordinates using the Bloch functions,
\begin{eqnarray*} 
V_{nm}(x) &=& \frac{e^2}{\kappa_{nm}}\int_0^{2\pi} \frac{d\varphi}{2\pi}
\frac{d\varphi'}{2\pi} \{ a^2 + (R_n-R_m)^2 \\ & +& x^2 
+ 4 R_n R_m \sin^2[(\varphi-\varphi')/2] \}^{-1/2} \;.
\end{eqnarray*}
The dielectric constants $\kappa_{nm}$ include the 
effect of the substrate and of insulating shells.
The Fourier transform $\widetilde{V}_{nm}(k)$ at long wavelengths
$|k R_N|\ll 1$ takes the form
\begin{equation} \label{vnm}
\widetilde{V}_{nm}(k) = -\frac{2e^2}{\kappa_{nm}} \ln (|k| \bar{R}_{nm})  \;,
\end{equation}
where the ``mean radius'' $\bar{R}_{nm}$ reads
\[
\ln \bar{R}_{nm} =
\int_0^{2\pi} \frac{d\varphi}{4\pi}\, \ln\{[(R_n-R_m)/2]^2+ R_n R_m \sin^2\varphi
\} \;.
\]
An order-of-magnitude estimate can be obtained by replacing
$\sin^2\varphi \to 1/2$, yielding
$\bar{R}_{nm} \approx  [ (R_n^2+R_m^2)/2 ]^{1/2}.$
The logarithmic singularity of $\widetilde{V}_{nm}(k\to 0)$ 
reflects the unscreened $1/r$ tail of the Coulomb interaction. 
In order to make contact with the usual LL concept, we restrict ourselves to a 
long-wavelength description by cutting off this singularity
at $k=2\pi/L$, where $L$ is the MWNT length \cite{bellucci99}.  
Including a factor $4/\pi v$ for later convenience, this leads to 
the dimensionless coupling constants 
\begin{equation} \label{coupl}
U_{nm}= \frac{8e^2}{\pi v \kappa_{nm} } \ln (L/2\pi \bar{R}_{nm}) \;.
\end{equation}
For $L=1~\mu$m, $\bar{R}_{nm}=6$~nm and $\kappa_{nm}=1.4$, 
Eq.~(\ref{coupl}) yields $U_{nm}\approx 16$.
Therefore both the intra- and the inter-shell electrostatic interactions 
are typically very strong and of comparable magnitude.  Due to the
weak logarithmic dependence of $U_{nm}$ on $\bar{R}_{nm}$,  
the approximation $U_{nm}=U$ can already yield sensible results. 

By virtue of bosonization, $H_0+H_I$ can now be diagonalized.
The bosonized form of the 1D fermion operators is \cite{book}
\begin{eqnarray}\nonumber
\psi_{r\alpha\sigma n}(x)&=& \frac{\eta_{r\alpha\sigma n}}
{\sqrt{2\pi a}} \exp\Bigl\{ i q^{}_F r x + i (\pi/4)^{1/2} 
[\theta_{c+,n}+r\varphi_{c+,n}   \\ \nonumber
&+& \alpha \theta_{c-,n} + r\alpha \varphi_{c-,n}
+\sigma \theta_{s+,n}+r\sigma\varphi_{s+,n} \\ \label{bos}
&+&\alpha\sigma \theta_{s-,n}+r\alpha\sigma\varphi_{s-,n}]
\Bigr\} \;,
\end{eqnarray}
where symmetric and antisymmetric 
linear combinations of the two transport bands $\alpha=\pm$ 
have been formed for charge  and spin degrees of freedom. 
The resulting four channels
are labelled by the index $\gamma=(c+,c-,s+,s-)$. 
The boson phase fields obey the commutator algebra
\[
[\theta_{\gamma,n}(x),\varphi_{\gamma',n'}(x')] = -(i/2)\delta_{nn'}
\delta_{\gamma\gamma'} {\rm sgn}(x-x') \;,
\]
so that $\varphi_{\gamma,n}$ has  the canonical momentum 
$\Pi_{\gamma,n}=-\partial_x \theta_{\gamma,n}$. 
The Majorana fermions $\eta_{r\alpha\sigma n}$ ensure anticommutation
 relations between 1D fermions with different indices $r\alpha\sigma n$.
  Finally, $q_F=E_F/v$ is determined by external gate voltages.  
Due to inter-shell tunneling, 
the Fermi energy $E_F$ and hence $q_F$ are identical for all 
$N$ shells.  
Using Eq.~(\ref{bos}), the density is $\rho_n(x)= (4/\pi)^{1/2}
\partial_x \varphi_{c+,n}$, and as a consequence the Hamiltonian  
decouples in all four channels,  $H_0+H_I=\sum_\gamma H_\gamma$. 
The charged channel is described by 
\begin{eqnarray} \nonumber
H_{c+}&=&\frac{v}{2}\sum_{n,m=1}^N \int dx
\Bigl\{ \, [ \Pi_{c+,n}^2 + (\partial_x\varphi_{c+,n})^2 ]
\,\delta_{nm} \\  &+& U_{nm} \,\partial_x\varphi_{c+,n} 
\partial_x\varphi_{c+,m} \Bigr\} \;.
\label{cplus}
\end{eqnarray}
The three neutral channels correspond to Eq.~(\ref{cplus}) with $U_{nm}=0$. 
Diagonalizing $H_{c+}$ then 
leads to an eigenvalue problem similar to the one studied by Matveev
and Glazman \cite{matveev93} for many-channel quantum wires in semiconductor
heterostructures, 
\begin{equation}\label{eig}
\sum_{m=1}^N \left\{ (1-g_j^{-2}) \delta_{nm} + U_{nm} 
\right\} \Gamma_{mj} = 0 \;,
\end{equation}
where $\varphi_{c+,n}(x) = \sum_j \Gamma_{nj} \Phi_j(x)$ and 
$\Gamma_{nj}$ is an orthogonal matrix. In terms of the
new boson phase fields $\Phi_j$ and their canonically 
conjugate momenta $\widetilde{\Pi}_j$,
 $H_{c+}$ results in the standard LL form, 
\begin{equation} \label{hcplus}
H_{c+} = \frac{v}{2}\sum_{j=1}^N \int dx 
\left( \widetilde{\Pi}_j^2 + g_j^{-2} (\partial_x \Phi_j)^2
\right) \;.
\end{equation}
The LL interaction constants $g_j\leq 1$ for the $N$ eigenmodes described
by the phase fields $\Phi_j$ serve as a measure of the correlation 
strength \cite{book}. 

It is then easy to determine all scaling exponents of interest. 
The exponents $\eta_{b/e}$ of the tunneling density of states (TDOS), 
$\rho(E)\sim E^\eta$,  for tunneling of an 
electron into the outermost shell ($n=N$) in the bulk or close to the end 
of the tube are
\begin{eqnarray} \label{etab}
\eta_b &=& \frac{1}{8} \sum_{j=1}^N \Gamma_{Nj}^2 (g_j^{-1}
+ g_j^{} - 2 ) \;,\\ \label{etae}
\eta_e &=& \frac{1}{4} \sum_j \Gamma_{Nj}^2 (g_j^{-1} -1) \;.
\end{eqnarray}
These exponents govern the power laws $G\sim T^\eta$ of the 
temperatur-dependent linear conductance for tunneling into the MWNT 
and can be measured in the experimental setup of Ref.\cite{schoenen99},
where external leads contact only the outermost shell of the MWNT.  
Moreover, in the limit of weak disorder backscattering,
 the linear conductance corrections 
$\delta G \sim T^{-p}$ are characterized by the exponent
\begin{equation}
p= \frac12 \sum_j \Gamma_{Nj}^2 ( 1-g_j ) \;.
\end{equation}
At sufficiently low temperatures, disorder in a LL
always leads to a strong backscattering situation.
Then the conductance vanishes as $T^\beta$ for $T\to 0$, where $\beta=2\eta_e$.
Observation of the exponents $p$ and $\beta$ requires good contacts 
between external leads and the MWNT.

It follows from the above analysis that an
{\sl individual MWNT exhibits LL behavior on low energy scales}, 
where the LL exponents can be computed by solving the eigenvalue problem
(\ref{eig}). Of course, exactly the same reasoning 
applies to ropes of SWNTs, the only
difference being the precise form  of the couplings $U_{nm}$.
Indeed the logarithmic singularity in Eq.~(\ref{coupl})
and hence the order of magnitude
 of the couplings applies to both systems.
For $N=1$, all  expressions above reduce to the SWNT theory 
of Refs.~\cite{egger97,kane97}.

To make analytical progress, we now solve Eq.~(\ref{eig}) for the
special case where all
$U_{nm}$ are equal and given by $U$.  [In the estimates below,
we neglect the weak logarithmic $N$-dependence of Eq.~(\ref{coupl}) 
and put $U=16$.]  The eigenvalues are 
\begin{equation} \label{eigenvalues}
g_1 = g = (1+NU)^{-1/2} \;,\qquad g_{2,\ldots,N}=1 \;.
\end{equation}
The first value has eigenvector $\Gamma_{n1}=N^{-1/2}$ and is identified with
the standard collective LL plasmon mode, now characterized by a pronounced
$N$-dependence of the LL interaction parameter $g$. The remaining  
$N-1$ degenerate modes are not affected by the interactions
 and correspond to Fermi liquid quasiparticles.  From the orthogonality 
of $\Gamma_{nj}$, it follows that $\sum_{j>1} \Gamma_{Nj}^2 = 1-1/N$,
and the above exponents read
\begin{eqnarray}\label{etab1}
\eta_b &=& \frac{1}{8N}\left( \frac{1}{g}+g-2\right) \;, \\ \label{etae1}
\eta_e &=& \frac{1}{4N} \left( \frac{1}{g}-1\right) = \beta/2 \;, \\
p &=& \frac{1}{2N}(1-g) \;.
\end{eqnarray}
For $N\to\infty$, the exponents $\eta_{b,e}$ and 
$\beta$ approach zero (the Fermi liquid value)
 only as $N^{-1/2}$, because the LL parameter $g$ also goes to zero
as $N^{-1/2}$.  This slow approach implies that LL power
laws should be well observable in typical MWNTs by focusing on the
TDOS. 
 For instance, for $N=10$ the exponents are 
$\eta_e=0.292$ and $\eta_b=0.135$,  while the
corresponding SWNT values are $\eta_e=0.780$ and
$\eta_b=0.295$.  On the other hand, the
exponent $p$ scales as $1/N$ for large $N$ and will thus
vanish more rapidly. 
 For instance, for $N=10$ it is $p=0.046$,
but for $N=1$ we would get $p=0.379$.   
LL power laws in MWNTs or ropes of SWNTs are thus generally much more pronounced 
for the TDOS than for the backscattering corrections in the presence of weak
disorder.  This important fact may be used in practice to obtain information about
the number $N$ of conducting shells.

So far single-electron hopping between the shells has been ignored.
It can be written as
\begin{equation} \label{hopp}
H_{t} = \sum_{n,m=1}^N T_{nm} 
\int dx \,\sum_{r\alpha\sigma}
 \psi^\dagger_{r\alpha\sigma n} \psi^{}_{r\alpha\sigma m} 
\;,
\end{equation}
where the hopping matrix acts only between nearest-neighbor shells, 
$T_{n m}= -t_n\delta_{n+1,m}- t_{n-1}\delta_{n,m+1}$.
The hybridization matrix elements are
of the order $t_n\approx 250$~meV \cite{schoenen99}. 
 Processes where $r\alpha \sigma$-type fermions
are scattered into different $r\alpha\sigma$-states 
are suppressed against Eq.~(\ref{hopp}) by momentum conservation and by
 higher scaling dimensions.  Renormalization group arguments suggest that
$H_t$ is relevant, so one has to be careful about its effect.

To include the hopping, we go back and first diagonalize $H_0+H_t$.
With the orthogonal matrix $Q_{n\nu}$, 
\begin{equation}\label{rotate}
\psi_{r\alpha\sigma n}(x) = \sum_{\nu=1}^N Q_{n\nu} 
\widetilde{\psi}_{r\alpha\sigma \nu}(x) \;,
\end{equation}
the rotated fermions $\widetilde{\psi}_{r\alpha\sigma \nu}$  
again obey the 1D Dirac Hamiltonian (\ref{h0}).
Denoting the $N$ eigenvalues of $T_{nm}$ as $T_\nu$, the respective 
eigenvectors span $Q_{n\nu}$ so that
$H_t = \sum_\nu  T_\nu \int dx\, \widetilde{\rho}_\nu(x)$,
with 1D densities $\widetilde{\rho}_\nu$ defined in analogy to 
Eq.~(\ref{d1d}).  In the long-wavelength limit, Eq.~(\ref{hi}) now reads
\begin{eqnarray}\label{four}
H_I &=& \frac{v}{2} \sum_{\nu_1\nu_2\nu_3\nu_4} Y_{\nu_1\nu_2\nu_3\nu_4} 
\sum_{r\alpha\sigma, r'\alpha'\sigma'} \\ &\times& \nonumber \int dx\, 
\widetilde{\psi}^\dagger_{r\alpha\sigma \nu_1}
\widetilde{\psi}^{}_{r\alpha\sigma\nu_2}
\widetilde{\psi}^\dagger_{r'\alpha'\sigma' \nu_3}
\widetilde{\psi}^{}_{r'\alpha'\sigma' \nu_4} \;,
\end{eqnarray}
with matrix elements
\begin{equation}\label{ys}
Y_{\nu_1\nu_2\nu_3\nu_4} = \sum_{nm} U_{nm} Q_{\nu_1 n}
 Q_{n\nu_2} Q_{\nu_3 m} Q_{m \nu_4} \;.
\end{equation}
The complicated four-fermion interactions (\ref{four})
are to a large degree responsible for
the technical difficulties encountered in previous studies of coupled LLs 
\cite{book}. 
According to Eq.~(\ref{coupl}), however, in MWNTs the couplings $U_{nm}$ 
are approximately equal, which allows to make further progress.
The situation might be different in ropes of SWNTs, 
where one should first diagonalize
$H_0+H_I$ by solving Eq.~(\ref{eig}),  and only afterwards 
include $H_t$, e.g., by perturbation theory. 

In MWNTs, the dominant matrix elements in Eq.~(\ref{ys})
are $Y_{\nu\nu\nu'\nu'}\equiv W_{\nu\nu'}=W_{\nu'\nu}$.  In fact, if $U_{nm}=U$,
 orthogonality of $Q_{n \nu}$ implies that $W_{\nu\nu'}=U$ and all other
matrix elements in Eq.~(\ref{ys}) are zero.  Keeping only the
couplings $W_{\nu\nu'}$,  Eq.~(\ref{four}) simplifies to
\begin{equation} \label{hii}
H_I= \frac{v}{2} \sum_{\nu\nu'}  W_{\nu\nu'}
\int dx \, \widetilde{\rho}_\nu(x) \widetilde{\rho}_{\nu'}(x)\;.
\end{equation}
Remarkably, the model $H_0+H_I+H_t$ can then be solved exactly
by bosonization in the rotated basis.
Using Eq.~(\ref{bos}) for $\widetilde{\psi}_{r\alpha\sigma \nu}(x)$, 
the neutral channels are described by the same Hamiltonian  as before. The 
only change arises in the charged sector, 
\begin{equation} \label{hcplusnew}
H_{c+} = \frac{v}{2}\sum_{j=1}^N \int dx 
\left( \widetilde{\Pi}_j^2 + g_j^{-2} (\partial_x \Phi_j)^2
\right)  + \sum_j \epsilon_j N_j \;,
\end{equation}
where $\epsilon_j = \sum_\nu T_\nu \Gamma_{\nu j}$. The ``zero modes'' $N_j= 
(4/\pi)^{1/2} \int dx \, \partial_x \Phi_j$ denote the total
number of particles in the $j$th eigenmode. 
The eigenvalues $g_j$ and the matrix $\Gamma_{\nu j}$ are then determined from
Eq.~(\ref{eig}) with $U_{nm}$ being replaced by $W_{\nu\nu'}$.
For $U_{nm}=U$, the only yet important effect of the
hopping consists of a splitting of the $N$ 1D transport
bands, with all plasmon excitations and LL interaction
constants $g_j$ being unaffected.
Using $\sum_j \Gamma_{\nu j} \epsilon_j=T_\nu$,
the TDOS for tunneling into the outermost shell is generally of the form
\begin{equation}
\rho(E) = \sum_\nu Q^{}_{N\nu} Q^\dagger_{N\nu} \; \rho_\nu(E-T_\nu) \;,
\end{equation}
where $\rho_\nu(E) \sim E^{\eta_{\nu,b/e}}$. The bulk/end exponents $\eta_{\nu,b/e}$ 
for tunneling into the $\nu$th MWNT eigenmode   
are given by Eqs.~(\ref{etab}) and (\ref{etae}), respectively,
with the replacement $\Gamma_{Nj}\to \Gamma_{\nu j}$. 
For $U_{nm}=U$, both exponents are independent of $\nu$
and given by Eqs.~(\ref{etab1}) and (\ref{etae1}).

Next the other interactions in Eq.~(\ref{four}) are briefly analyzed.
If three or four indices $\nu_i$ are different, the resulting 
perturbations are always highly irrelevant since single-particle 
excitations for more than two modes are involved in the scattering. 
The contribution with $\nu_1=\nu_3$ and $\nu_2=\nu_4$ 
corresponds to a simultaneous transfer of two fermions from one
mode to another, which is also an irrelevant process.
The remaining interactions are due to the couplings
$B_{\nu \nu'} = B_{\nu' \nu}= Y_{\nu\nu'\nu'\nu}$, which
are small for $U_{nm}\approx U$.
Using bosonization, one observes that the $B_{\nu\nu'}$ are 
{\sl marginally relevant}\, scaling fields 
causing a weak renormalization of the LL interaction
parameters also in the neutral channels [which were all equal
to one before], and the nonlinear contribution 
\[
H_I^\prime = \frac{4v}{(\pi a)^2} \sum_{\nu<\nu'} B_{\nu\nu'} \int dx \, 
 \prod_\gamma \cos\{\sqrt{\pi} [\varphi_{\gamma,\nu}(x)-\varphi_{\gamma,\nu'}(x)] 
\}  \;.
\]
Since marginally relevant scaling fields lead to exponentially
small gaps for small couplings, the generic behavior of MWNTs is 
indeed described in terms of the LL model outlined above.
The predicted pseudogap in ropes \cite{delaney98} can then be 
rationalized by noting that the $B_{\nu\nu'}$ are 
presumably larger in ropes.

How will  Fermi liquid behavior be restored as the number of conducting
shells approaches infinity?
Taking for simplicity all nearest-neighbor hybridization matrix elements $t_n=t$, 
the eigenvalues of the hopping matrix for large $N$ 
read $T_\nu= -2t \cos(2\pi \nu/N)$, with $Q_{N\nu} Q^\dagger_{N\nu}=1/N$. 
Hence the  TDOS becomes 
\begin{equation}
\rho(E)\sim N^{-1} \sum_\nu (E-T_\nu)^{\eta_{b/e}} \;,
\end{equation}
where the threshold energies $T_\nu$ become narrowly spaced.
In effect, for large enough $N$, the LL singularities will then be smeared out
simply because subsequent thresholds are too close together.
The approach to the
Fermi liquid thus proceeds by (i) a $N^{-1/2}$ 
decrease in the exponents $\eta_{b/e}$, and (ii) tunneling thresholds 
that become closer and closer spaced.  
For moderate values of $N$, however, different 
$T_\nu$ are sufficiently well separated to allow for
the observation of LL power laws.

To conclude, the low-energy theory of an individual MWNT on 
an insulating substrate has been given.  Including the long-ranged
Coulomb interaction,  pronounced correlation effects are predicted
which can be understood in the framework of the Luttinger liquid model.
Clearly, there are many open questions to be addressed in future work,
e.g., the effects of a parallel or perpendicular magnetic field,
or the virtual or real population of higher subbands as the 
energy scale increases. 

I wish to thank L.~Forr{\'o}, A.~Gogolin, H.~Grabert, and C.~Sch{\"o}nenberger
for helpful discussions and acknowledge 
support by the Deutsche Forschungsgemeinschaft
(Bonn) under the Gerhard-Hess program.


\begin{references}
\bibitem{foot0}  Since statistically 2/3 of all tubes are metallic,
typically $N\approx 10$.
\end{references}
\end{document}